\documentclass[aps,prl,twocolumn,tightenlines,superscriptaddress]{revtex4}
\usepackage{epsfig,rotating}
\usepackage{multirow}
\newcommand{\nuc}[2]{\hbox{$^{#1}$#2}}

\begin{document}
%\draft
\title{Is the structure of \nuc{42}{Si} understood?}

\author{A.\ Gade}
   \affiliation{National Superconducting Cyclotron Laboratory,
      Michigan State University, East Lansing, Michigan 48824, USA}
   \affiliation{Department of Physics and Astronomy,
      Michigan State University, East Lansing, Michigan 48824, USA}
\author{B.\,A.\ Brown}
    \affiliation{National Superconducting Cyclotron Laboratory,
      Michigan State University, East Lansing, Michigan 48824, USA}
   \affiliation{Department of Physics and Astronomy,
      Michigan State University, East Lansing, Michigan 48824, USA}
\author{J.\ A.\ Tostevin}
     \affiliation{Department of Physics, University of Surrey,
       Guildford, Surrey GU2 7XH, United Kingdom}
\author{D.\ Bazin}
    \affiliation{National Superconducting Cyclotron Laboratory,
      Michigan State University, East Lansing, Michigan 48824, USA}
    \affiliation{Department of Physics and Astronomy,
      Michigan State University, East Lansing, Michigan 48824, USA}
\author{P.\ C.\ Bender}
     \altaffiliation{Present address: Department of Physics, University of Massachusetts Lowell, Lowell, Massachusetts 01854, USA}
    \affiliation{National Superconducting Cyclotron Laboratory,
      Michigan State University, East Lansing, Michigan 48824, USA}
\author{C.\,M.~Campbell}
      \affiliation{Nuclear Science Division, Lawrence Berkeley
          National Laboratory, California 94720, USA}
\author{H.\ L.\ Crawford}
        \affiliation{Nuclear Science Division, Lawrence Berkeley
          National Laboratory, California 94720, USA}
\author{B.\ Elman}
    \affiliation{National Superconducting Cyclotron Laboratory,
      Michigan State University, East Lansing, Michigan 48824, USA}
    \affiliation{Department of Physics and Astronomy,
      Michigan State University, East Lansing, Michigan 48824, USA}
\author{K.\ W.\ Kemper}
    \affiliation{Department of Physics,
         Florida State University, Tallahassee, Florida 32306, USA}
\author{B.\  Longfellow}
    \affiliation{National Superconducting Cyclotron Laboratory,
      Michigan State University, East Lansing, Michigan 48824, USA}
    \affiliation{Department of Physics and Astronomy,
      Michigan State University, East Lansing, Michigan 48824, USA}
\author{E.\ Lunderberg}
    \affiliation{National Superconducting Cyclotron Laboratory,
      Michigan State University, East Lansing, Michigan 48824, USA}
    \affiliation{Department of Physics and Astronomy,
      Michigan State University, East Lansing, Michigan 48824, USA}
%\author{A.\ Poves}
%      \affiliation{Departamento de F\'{i}sica Te\'{o}rica e IFT-UAM/CSIC,
%        Universidad Aut\'{o}noma de Madrid, E-28049 Madrid, Spain}
\author{D.\ Rhodes}
    \affiliation{National Superconducting Cyclotron Laboratory,
      Michigan State University, East Lansing, Michigan 48824, USA}
    \affiliation{Department of Physics and Astronomy,
      Michigan State University, East Lansing, Michigan 48824, USA}
%\author{L.\ A.\ Riley}
%      \affiliation{Department of Physics and Astronomy, Ursinus College,
%        Collegeville, Pennsylvania 19426, USA}
\author{D.\ Weisshaar}
    \affiliation{National Superconducting Cyclotron Laboratory,
      Michigan State University, East Lansing, Michigan 48824, USA}
\date{\today}

\begin{abstract}
A more detailed test of the implementation of nuclear forces that drive 
shell evolution in the pivotal nucleus \nuc{42}{Si} -- going beyond earlier
comparisons of excited-state energies -- is important. The two leading
shell-model effective interactions, SDPF-MU and SDPF-U-Si, both of which
reproduce the low-lying \nuc{42}{Si}($2^+_1$) energy, but whose predictions for
other observables differ significantly, are interrogated by the population of
states in neutron-rich \nuc{42}{Si} with a one-proton removal 
reaction from \nuc{43}{P} projectiles at 81~MeV/nucleon. The measured cross 
sections to the individual \nuc{42}{Si} final states are compared to
calculations that combine eikonal reaction dynamics with these shell-model
nuclear structure overlaps. The differences in the two shell-model descriptions
are examined and linked to predicted low-lying excited $0^+$ states and shape
coexistence. Based on the present data, which are in better agreement with the
SDPF-MU calculations, the state observed at 2150(13)~keV in \nuc{42}{Si} is
proposed to be the ($0^+_2$) level.  
\end{abstract}

\maketitle

Modeling the nuclear landscape with predictive power, including 
the most exotic nuclei near the limits of nuclear existence, is 
an overarching goal driving 21$^{st}$ century nuclear science. This 
quest thrives through the interplay of experiment and theory, 
whereby observables measured for very neutron-proton asymmetric 
nuclei reveal isospin-dependent aspects of the nuclear force. 
They also identify benchmark nuclei, critical for understanding 
and for quantitative extrapolations toward the shortest-lived 
rare isotopes -- many outside of the reach of laboratory studies 
but whose properties underpin the modeling of nucleosynthesis 
processes, for example. Over the few decades of rare-isotope 
research, certain nuclei defying textbook expectations have emerged as pivotal
--  they are typically located in regions of rapid structural change or at the 
extremes of weak binding where open quantum systems properties
are exhibited. The $Z=14$ isotope \nuc{42}{Si}$_{28}$ is one 
such nucleus.

At present, the most neutron-rich Si isotope known to exist 
is \nuc{44}{Si}, with neutron number $N=30$~\cite{Tar07}, and 
the most neutron-rich $N=28$ isotone with known spectroscopic 
information is \nuc{40}{Mg}~\cite{Cra19}. This places their 
even-even neighbor \nuc{42}{Si} ($Z=14,N=28$) at the frontier 
of nuclear experimentation. A description of \nuc{42}{Si} has
challenged nuclear structure physics for a long time. Early on, 
the $\beta$-decay half-life of \nuc{42}{Si}~\cite{Gre04} and 
the particle stability of \nuc{43}{Si}~\cite{Not02} were 
interpreted as indicators that the $N=28$ magic number had 
broken-down, but that a pronounced $Z=14$ sub-shell closure 
may prevent \nuc{42}{Si} from being well
deformed~\cite{Cot98,Cot02,Fri05,Fri06}. These speculations were resolved by the  
first successful spectroscopy of \nuc{42}{Si}~\cite{Bas07}, 
revealing a surprisingly low-lying first $2^+$ state, at 
$E(2^+_1) = 770(19)$~keV, the onset of collectivity, and 
the breakdown of the $N=28$ magic number in \nuc{42}{Si}.

Reproducing this evolution, (a) along the Si isotopic chain, 
starting from doubly-magic \nuc{34}{Si}$_{20}$, with the 
rapid increase in collectivity or deformation at $N=28$,
and (b) along the isotone line from doubly-magic \nuc{48}
{Ca}$_{28}$ towards Si, has been a formidable challenge 
for the nuclear shell model. Two shell-model effective 
interactions, SDPF-U~\cite{Now09} and SDPF-MU~\cite{Uts12}, 
succeeded to reproduce a low-lying $2^+_1$ state in \nuc{42}
{Si}~\cite{footnote}. The mechanism underlying the collapse of the $N=28$ 
shell gap was attributed to: (i) the filling of the neutron 
$0f_{7/2}$ orbit reducing the $Z=14$ gap relative to \nuc{34}{Si}, and, in concert, (ii) the removal of protons from the 
$0d_{3/2}$ orbit reducing the $N=28$ gap relative to \nuc{48}{Ca}, both the
result of the proton-neutron monopole parts  
of the tensor force~\cite{Ots13}. $\Delta l,j =2$ quadrupole 
correlations, reaching across the so-narrowed $Z=14$ and 
$N=28$ gaps, then mutually enhance one another leading to 
deformation, as argued within the context of an SU(3)-like 
scheme~\cite{Bas07,Now09} or a nuclear Jahn-Teller effect~
\cite{Uts12}. While both shell-model interactions reproduce 
the low energy first-excited \nuc{42}{Si}($2^+$) state, 
their predictions for the level density and energies of states
beyond the first $2^+$ differ dramatically. This demands
confrontation with additional experimental data to validate
these different implementations of the suspected drivers
of rapid shell evolution in this benchmark
region~\cite{Rod02,Gau10,Sor13,Cau14,Gad16a,Ots19}, where the spectrum of
the near-dripline nucleus \nuc{40}{Mg} turned out to be surprising~\cite{Cra19}. 

It required half a decade and a new-generation accelerator
facility for spectroscopy beyond the \nuc{42}{Si} first
excited state to be performed~\cite{Tak12}. There, the
\nuc{12}{C}(\nuc{44}{S},\nuc{42}{Si}$+\gamma$)X two-proton
removal reaction was used to populate excited states in
\nuc{42}{Si}. The first $4^+$ state was suggested at 2173(14)~keV, with the
ratio $R_{4/2} = E(4^+_1)/E(2^+_1)$ 
close to the rotational limit, as one may expect for a
well-deformed nucleus~\cite{Tak12}. However, a direct reaction model analysis, using the SDPF-U/SDPF-U-Si 
and SDPF-MU shell-model two-nucleon amplitudes~\cite{Tos13}, 
could not reconcile the $\gamma$-ray spectra and assignments
reported in~\cite{Tak12}, indicating that \nuc{42}{Si}
was not understood within the current shell-model picture
after all; one-proton removal to \nuc{42}{Si} was proposed to clarify the situation~\cite{Tos13}. 

Here, we report this first high-resolution in-beam $\gamma$-ray
spectroscopy of \nuc{42}{Si} in the direct one-proton removal
reaction \nuc{9}{Be}(\nuc{43}{P},\nuc{42}{Si}$+\gamma$) using
GRETINA~\cite{gretina,Wei17}. The measured partial removal cross
sections are compared to direct reaction calculations combining
eikonal dynamics and shell-model spectroscopic factors. We probe 
the different implementations of the drivers of shell evolution 
on the valence single-particle levels through the theoretical 
spectroscopic factors from the SDPF-MU and SDPF-U-Si
shell-model calculations. The stark differences in observables
(other than the $2^+_1$ energy) predicted by the two shell-model
descriptions of \nuc{42}{Si} reveal that this key nucleus is
not yet sufficiently understood.

The secondary beam of \nuc{43}{P} was produced by fragmentation
of a 140 MeV/u stable \nuc{48}{Ca} beam, delivered by the Coupled
Cyclotron Facility at NSCL~\cite{Gad16b}, impinging on a 1363~mg/cm$^2$
\nuc{9}{Be} production target and separated using a 
150~mg/cm$^2$ Al degrader in the A1900 fragment separator~
\cite{a1900}. The momentum acceptance of the separator was set
to transmit $\Delta p/p=3$\%, yielding rates of typically 45
\nuc{43}{P}/second. About 20\% of the secondary beam composition was
\nuc{43}{P}, with \nuc{42}{P} and \nuc{44}{S} as the most
intense other components.

The secondary \nuc{9}{Be} reaction target (476~mg/cm$^2$ thick)
was located at the target position of the S800 spectrograph.
Reaction products were identified on an event-by-event basis
in the S800 focal plane with the standard focal-plane detector
systems~\cite{s800}. The inclusive cross section for the one-proton
knockout from \nuc{43}{P} to \nuc{42}{Si} was measured to be 
$\sigma_{inc}=3.4(2)$~mb.

\begin{figure}[h]
\epsfxsize 8.2cm
\epsfbox{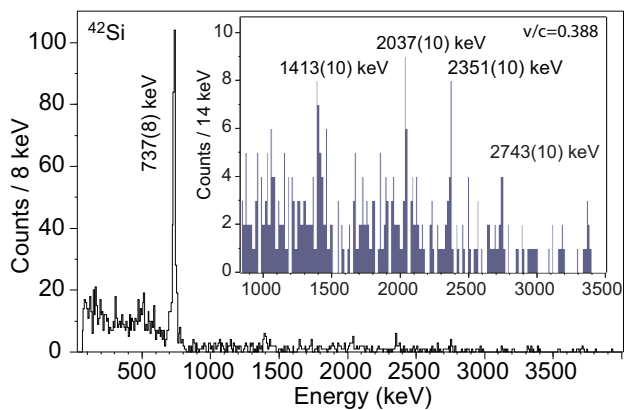}
\caption{\label{fig:gamma} Gamma-ray
spectrum in coincidence with \nuc{42}{Si} reaction residues,
event-by-event Doppler-reconstructed and including nearest-neighbor
addback. The inset shows the high-energy region expanded.}
\end{figure}

The $\gamma$-ray detection system GRETINA~\cite{gretina,Wei17},
an array of 40 high-purity Ge crystals 
that are each 36-fold segmented, was used to detect the
prompt $\gamma$ rays emitted by the reaction residues. The
ten detector modules -- with four crystals each -- were
arranged in two rings, with four modules located at
58$^{\circ}$ and six at 90$^{\circ}$ with respect to the
beam axis. Online signal decomposition provided $\gamma$-ray
interaction points ($xyz$) for event-by-event Doppler
reconstruction of the photons emitted in-flight at $v/c
\approx 0.4$. The information on the momentum vector of
projectile-like reaction residues, as reconstructed through
the spectrograph, was incorporated into the Doppler correction.
Figure~\ref{fig:gamma} shows the Doppler-reconstructed
$\gamma$-ray spectrum for \nuc{42}{Si} with nearest-neighbor
addback included~\cite{Wei17}. It is apparent that only
little cross section is carried by excited states beyond
the $2^+_1$ level. Nevertheless, the remarkable peak-to-background
ratio allows for spectroscopy at such modest levels of statistics
and, as shown in the inset of Fig.~\ref{fig:gamma}, weak peak
structures at 1413(10), 2037(10), 2351(10), and
2743(10)~keV are visible, in addition to the strong $2^+_1
\rightarrow 0^+_1$ transition at 737(8)~keV. The lowest three
of these higher-energy $\gamma$ rays likely correspond to the 1431(11), 2032(9),
and 2357(15)~keV transitions reported in~\cite{Tak12}.

\begin{figure}[h]
\epsfxsize 8.4cm
\epsfbox{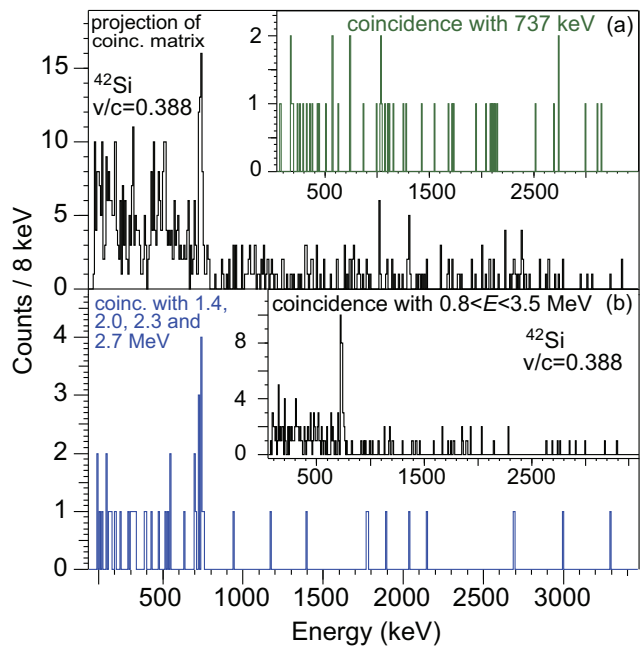}
\caption{\label{fig:coinc} (a) Projection of the $\gamma\gamma$
  coincidence matrix and coincidences with 737~keV (inset). 
(b) Summed coincidence spectrum obtained from gates on
the photopeaks of all higher-energy $\gamma$-ray transitions and
coincidences with $0.8 < E < 3.5$~MeV (inset). Spectra are not background-subtracted.}
\end{figure}

In spite of the low statistics at high excitation energy, a
coincidence analysis provides some limited guidance for the
placement of the transitions in the level scheme. Figure~\ref{fig:coinc}(a) shows the projection of 
the $\gamma \gamma$ coincidence matrix and the coincidences
with the $2^+_1 \rightarrow 0^+_1$ transition (inset). In
comparison to the $\gamma$-ray singles spectrum of Fig.~\ref{fig:gamma}, the
projection of the coincidence matrix 
shows a significantly increased number of counts at $E >
800$~keV relative to the 737-keV peak counts, indicating
that the high-energy region bears coincidences. Due to the
low statistics, no peaks are expected in the coincidence
spectrum (inset) but groups of counts appear to cluster where, with
more statistics, the peaks and/or Compton edges of the
transitions reported here would occur. Turning the
analysis around and showing the sum of cut spectra coincident
with the 1.4, 2.0, 2.3 and 2.7~MeV photopeaks returns the
$2^+_1 \rightarrow 0^+_1$ transition at about the right
intensity for all higher-lying transitions to be coincident
with it. The inset shows a coincident spectrum to the broad
energy region of $0.8 < E < 3.5$, now including, in addition
to the photopeaks, also the Compton continua. The number
of counts in the $2^+_1\rightarrow 0^+_1$ is increased by
a factor of about three as one would expect from the
peak-to-Compton ratio of GRETINA at these energies. We, therefore,
tentatively propose that all of the higher-lying transitions
reported here feed the first $2^+$ state. All the resulting
excited states lie below the (rather uncertain) neutron
separation energy of $S_n=3721(747)$~keV.

The photopeak efficiency of GRETINA was calibrated with standard sources and
corrected for the Lorentz boost of the $\gamma$-ray distribution emitted by the
residual nuclei moving at almost 40\% of the speed of light and addback factors
from GEANT simulations~\cite{geant}. Partial cross sections to the specific
final states were determined from the efficiency-corrected $\gamma$-ray peak
areas, with discrete feeding subtracted, relative to the number of incoming
\nuc{43}{P} projectiles and the number density of the target.

One-nucleon removal is a direct reaction with sensitivity to single-particle 
degrees of freedom. The cross sections for the population of individual
states in the reaction residue depend sensitively on the overlap, and 
spectroscopic factor, of the projectile initial and the residue final 
states~\cite{knock}. The shape of the ground-state residue parallel momentum
distribution in the one-proton removal from \nuc{44}{S} to \nuc{43}{P}
unambiguously revealed
the knockout of an $s_{1/2}$ proton, determining the ground-state 
spin of \nuc{43}{P} to $1/2^+$~\cite{Ril08}, in agreement with shell model.

\begin{figure}[h]
\epsfxsize 7.8cm
\epsfbox{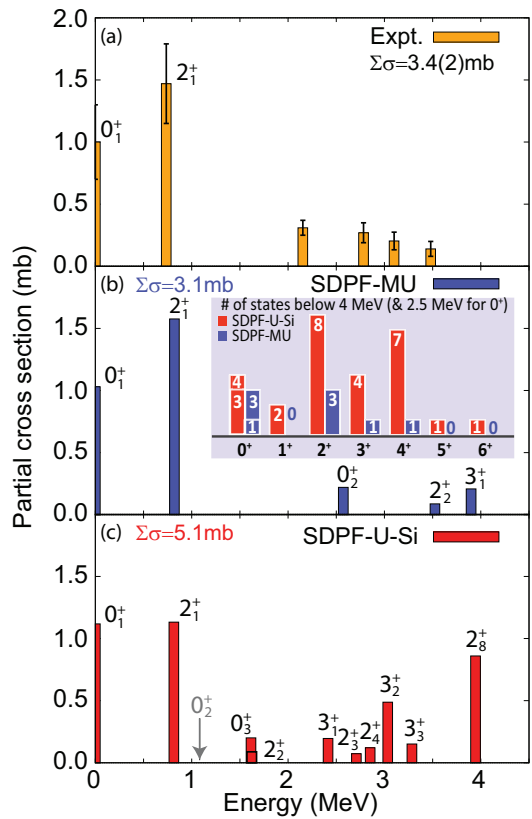}
\caption{\label{fig:xsec} Partial proton removal cross sections from \nuc{43}{P}
to bound states in \nuc{42}{Si}: (a) experiment, (b) direct reaction theory with
SDPF-MU shell-model spectroscopic factors, and (c) direct reaction theory with
SDPF-U-Si shell-model spectroscopic factors. For the calculations, states up
to 4~MeV and carrying $C^2S > 0.02$ spectroscopic strengths were included and
$R_s=0.3$ was applied. The 
inset to (b) shows, 
for each of the effective interactions, the number of states below 4~MeV (and
additionally below 2.5~MeV for the $0^+$ states).}
\end{figure}

Using the one-nucleon removal reaction methodology of Ref. \cite{Gad08} and shell-model spectroscopic factors, the partial 
cross sections to all bound, shell-model \nuc{42}{Si} final states were
calculated. These are confronted with experiment in Fig.~\ref{fig:xsec}. A
reduction factor $R_s=0.3$, appropriate for the effective proton-neutron 
separation energy asymmetry from \nuc{43}{P}, $\Delta S \approx$16~MeV
\cite{AME2016,Tos14}, is applied to the calculated cross sections. The $R_s$ and $\Delta S$ deduced from the measured and calculated cross sections (using SDPF-MU) are 0.33(2) and 15.6~MeV.

The measured cross-section distribution reflects the rather simple
$\gamma$-ray spectrum, dominated by the $2^+_1 \rightarrow 0^+_1$
transition, with weak higher-energy transitions. The majority of the
cross section feeds the ground state and the $2^+_1$ level, with modest
spectroscopic strength distributed between 2 and 3.5~MeV. The partial
cross sections calculated with the SDPF-MU spectroscopic factors describe
the measured cross section distribution well, including the values of 
$\sigma_{inc}$, $\sigma(0^+_1)$ and $\sigma(2^+_1)$ on the absolute scale
and the fraction of the strength at higher excitation energy. Use of the
SDPF-U-Si wavefunctions predicts a larger inclusive cross section and
significantly more strength above 1.5~MeV, in particular, if the 
predicted $2^+_8$ state at 3.945~MeV were bound. The cross section 
distribution based on the SDPF-MU spectroscopy also better matches the
measured distribution on a detailed level. The states predicted to
be populated strongly are calculated to decay predominantly to the first $2^+$ state, consistent with our proposed level scheme. The larger
strength at higher excitation energy, predicted using SDPF-U-Si, is
not supported by the $\gamma$-ray spectrum (see Fig.~\ref{fig:gamma}).
For example, the $3^+_2$ state at 3.034~MeV, predicted to carry
significant strength, would decay with a $>90$\% branch to the $3^+_1$
state with a $\sim$600~keV $\gamma$-ray transition that should be visible
in the data with $\sim$60 peak counts. Similarly, if the $2^+_8$ state were
bound, the measured 
inclusive cross section should have been 30\% higher and a $\gamma$-ray
transition of order five times stronger than the 2.7~MeV peak should
have been observed near 3.2~MeV. We conclude that the SDPF-MU
interaction provides calculations in better agreement with the data
than SDPF-U-Si. 

This outcome seems rooted in the vastly different
\nuc{42}{Si} level densities predicted using SDPF-U-Si and SDPF-MU.
The insert to Fig.~\ref{fig:xsec}(b) illustrates this point through
the number of states per $J^+$ value below 4~MeV (and also below
2.5~MeV for $0^+$ states). SDPF-U-Si offers five more 2$^+$ and
three more 3$^+$ states in this energy window, some of them predicted to carry
substantial spectroscopic strengths and thus proton-removal cross section.

\begin{figure}[h]
\epsfxsize 8.4cm
\epsfbox{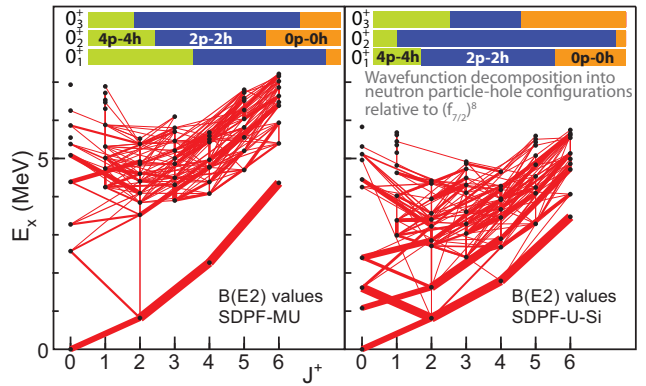}
\caption{\label{fig:mess} Wavefunction decomposition of the first three $0^+$
  states and network of $E2$ transitions as predicted by the
SDPF-MU (left) and SDPF-U-Si (right) shell-model calculations. The first
ten states of each $J^+$ are computed together with their $B(E2)$ strength 
connecting to all other calculated states (displayed as connecting lines).
The line thicknesses represent the $B(E2)$ strength of each transition.}
\end{figure}

Perhaps the most remarkable difference is the number of low-lying
$0^+$ states generated by the two shell-model interactions, namely
4(3) and 3(1) below 4(2.5)~MeV (including the ground state), from SDPF-U-Si and
SDPF-MU, 
respectively. In fact, this abundance of low-lying $0^+$ states
in the SDPF-U-Si calculation appears to drive the high density of
low-lying \nuc{42}{Si} levels, as compared to SDPF-MU. This is
illustrated in Fig.~\ref{fig:mess} where, for the first ten
calculated states for each  $J^+$ quantum number, the predicted
$B(E2)$ electric quadrupole transition strengths to all other
levels are indicated by lines. Here the line thickness scales
with the $B(E2)$ values. Both calculations show a pronounced
yrast line, formed by the strong intraband $E2$ decays between
the first states of each even-$J$ spin. For SDPF-MU, the $0^+_2$
to $0^+_4$ states are located beyond 2.5~MeV in excitation energy
and are weakly connected with $E2$ transitions to the higher-lying
$2^+$ states that occur with significant level density above
3-4~MeV. The (isomeric) excited $0^+_2$ state within SDPF-U-Si,
however, appears to be the band-head of an even-$J$ band that
carries collectivity comparable to the yrast band, as indicated
by the similar $B(E2)$ values. The third and fourth $0^+$ states
are then predicted to be strongly connected to higher-lying
$2^+$ states which appear with significant level density starting
at 2.5~MeV. The level structure from SDPF-U-Si is more compressed
than that from the SDPF-MU calculation, leading to the markedly
increased level density at low energies. The low-lying $0^+$
states within SDPF-U-Si seem to play a role in this, with the
second $0^+$ state and the band structure built on top, constituting
a remarkable case of predicted shape or configuration coexistence
with essentially no connecting $E2$ transitions to the yrast
band. Figure~\ref{fig:mess} also shows
the neutron particle-hole content of the three lowest-lying $0^+$ states
relative to the closed-shell configuration~\footnote{In this presentation, 2p-2h
  contains about 10\% of 1p-1h and 3p-3h content.}. Clearly, the wavefunctions of the
$0^+$ states differ significantly between the two calculations. Identifying and
characterizing the excited $0^+$ states and structures 
built on top of these will be a challenge for future experiments.   

Since \nuc{43}{P} has a $1/2^+$ ground state, only positive-parity states
up to and including $J^{\pi}=3^+$ can be populated directly by the removal
of an $sd$-shell proton (see also Fig.~\ref{fig:xsec}). So, if the 1413~keV 
$\gamma$ ray observed in this work corresponds to that reported in
~\cite{Tak12}, the tentative $(4^+_1)$ assignment made there for the
corresponding state is thus not tenable. In the SDPF-MU picture, which is
largely consistent with the present measurements, the possibility that
the 1413-keV transition is due to indirect feeding is rather unlikely,
since the populated $2^+_2$ and $3^+_1$ states are predicted to have
only minuscule decay branches, of around 0.6\% and 2\%, respectively,
to the $4^+_1$. From Fig.~\ref{fig:xsec} it seems, rather, that the
state at 2150(13)~keV may indeed be the first exited $0^+$ state,
consistent also with its cross section and excitation energy predicted
by SDPF-MU calculations. We note that the assignment of $0^+_2$ for the 
2150(13)~keV level was also most consistent with the (SDPF-MU) two-proton
removal cross section analysis presented in~\cite{Tos13}. One-proton
removal data from \nuc{43}{P} with sufficient statistics to examine
the shape of the parallel momentum distribution of \nuc{42}{Si} in
coincidence with the 1.4~MeV $\gamma$-ray transition would allow
confirmation of this assignment if an $\ell=0$ shape was found. At least an
order of magnitude more statistics would be needed. A similar analysis is also
possible for two-proton removal~\cite{Tos13}. This challenge may have to await 
future, high-statistics experiments at a new-generation facility.

In summary, high-resolution in-beam $\gamma$-ray spectroscopy with
GRETINA was performed for the neutron-rich nucleus \nuc{42}{Si} in
a one-proton removal reaction from \nuc{43}{P} projectiles. Five
$\gamma$-ray transitions are reported, four of which have been
observed previously. Coincidence data were used to propose a tentative
level scheme, which was then utilized to extract a partial cross section
distribution 
for the direct one-proton removal reaction. The measured partial cross
sections are confronted with direct reaction calculations that combine
eikonal reaction dynamics with SDPF-MU and SDPF-U-Si shell-model
spectroscopic information. These two effective interactions predict
markedly different low-lying level densities with the scenario painted
by the SDPF-MU calculations more consistent with the new data. This underscores
the difficulty in extrapolating  
configuration-interaction calculations towards the
neutron dripline and shows that nuclear models must be tested beyond
the energy of the lowest $2^+$ states. Our results highlight the SDPF-MU interaction as a starting point for understanding the role of weak
binding for the isotone \nuc{40}{Mg}, for which both shell-model effective interactions
fail to describe the observed, rather compressed, spectrum and where continuum
effects are suggested to be at play~\cite{Cra19}. From the 
selectivity of the reaction mechanism, and in agreement with similar
theoretical work on the two-proton removal reaction leading to \nuc{42}
{Si}, a level at 2150(13)~keV is proposed to be the $(0^+_2)$ state
rather than the previously suggested $(4^+_1)$ level. The differences
in calculations from the two shell-model effective interactions are 
discussed and the special role of the low-lying $0^+$ states is 
characterized. More final-state-exclusive experimental data are 
needed to further interrogate \nuc{42}{Si} and to clarify its 
description within the nuclear shell model. Ultimately, ab-initio-based
Hamiltonians that incorporate the effects of the continuum are needed.

\begin{acknowledgments}
This work was supported by the US National Science Foundation (NSF) 
under Cooperative Agreement No. PHY-1565546 and Grant No. PHY-1811855, by the US
Department of Energy (DOE) National Nuclear Security Administration through the
Nuclear Science and Security Consortium under award 
number DE-NA0003180, and by the DOE-SC Office of 
Nuclear Physics under Grant No. DE-FG02-08ER41556 (NSCL) and DE-AC02-05CH11231 (LBNL). GRETINA was 
funded by the DOE, Office of Science. Operation of the array at NSCL 
was supported by the DOE under Grant No. DE-SC0014537 (NSCL) and 
DE-AC02-05CH11231 (LBNL). J.A.T. acknowledges
support from the Science and Technology Facilities Council (U.K.) Grant
No. ST/L005743/1. Discussions with A. Poves are acknowledged.
\end{acknowledgments}


\begin{thebibliography}{10}
\bibitem{footnote} The effective interaction SDPF-U is formulated in two parts, one to be used for $Z \geq 15$ and one valid
  for $Z \leq 14$, called here SDPF-U-Si. This was taken into account for the
  overlaps used in this work.
\end{thebibliography}
\end{document}